\documentclass[aps,prd,twocolumn,groupedaddress,floatfix,nofootinbib,showpacs]{revtex4}
\usepackage{amsmath}
\usepackage{amsfonts}
\usepackage{amssymb}
\usepackage{latexsym}
\usepackage{graphicx}
\usepackage{bm}
\usepackage{epsfig}
\usepackage[english]{babel}

\begin{document}

%\preprint{\vbox{ \hbox{}   \hbox{} }}

\title{Coulomb Force as an Entropic Force}
\author{Tower Wang\footnote{Electronic address: wangtao218@pku.edu.cn}}
\affiliation{Center for High-Energy Physics, Peking University,\\
Beijing 100871, China\\ \vspace{0.2cm}}
\date{\today\\ \vspace{1cm}}
\begin{abstract}
Motivated by Verlinde's theory of entropic gravity, we give a
tentative explanation to Coulomb's law with an entropic force. When
trying to do this, we find the equipartition rule should be extended
to charges and the concept of temperature should be reinterpreted.
If one accepts the holographic principle as well as our
generalizations and reinterpretations, then Coulomb's law, the
Poisson equation and the Maxwell equations can be derived smoothly.
Our attempt can be regarded as a new way to unify the
electromagnetic force with gravity, from the entropic origin.
Possibly some of our postulates are related to the D-brane picture
of black hole thermodynamics.
\end{abstract}


\maketitle

%\tighten

%%%%%%%%%%%%%%%%%%%%%%%%%%%%%%%%%%%%%%%%%%

%\tableofcontents
\section{Motivation}\label{sect-mot}
Starting from the holographic principle and an equipartition rule,
E. Verlinde proposed an intriguing explanation for Newton's law of
gravity as an entropic force \cite{Verlinde:2010hp}. This idea turns
out to be very powerful.\footnote{As a partial list, some previous
and recent papers along this direction are
\cite{Jacobson:1995ab,Padmanabhan:2009kr,Caravelli:2010be,Li:2010cj}.}
From it one can derive both Newton's laws and Einstein's equations.
In this paper, we try to understand its implication to the
Reissner-Nordstr\"om (RN) spacetime. With a few generalizations and
reinterpretations, we find the same idea can produce Coulomb's law
and Maxwell's equations.

To make our narration clear, for the most part we will concentrate
on the 4-dimensional spacetime, in which the RN solution is
\begin{equation}\label{RN}
ds^2=-f(r)dt^2+\frac{1}{f(r)}dr^2+r^2d\Omega^2
\end{equation}
with
\begin{equation}
f(r)=1-\frac{2G_NM}{c^2r}+\frac{G_N^2Q^2}{c^4r^2},~~~~~M\geq|Q|.
\end{equation}
For conciseness, we work in the ``geometrized'' unit of charge so
that the Coulomb force between point charges $Q$, $q$ at large
separation $r$ in flat space takes the form
\begin{equation}\label{Coulomb}
F_{em}=\frac{G_NQq}{r^2}.
\end{equation}
If one feels uneasy with such a unit, the more familiar form can be
recovered by replacing $Q\rightarrow Q/\sqrt{4\pi\varepsilon_0G_N}$,
$q\rightarrow q/\sqrt{4\pi\varepsilon_0G_N}$ for electric charges,
where $\varepsilon_0$ is the electric constant and $G_N$ is the
Newton's constant.

Now let us show a puzzle regarding the equipartition rule. On the
one hand, in \cite{Verlinde:2010hp} it was proposed that the
temperature $T$ is determined by
\begin{equation}\label{equip0}
Mc^2=\frac{1}{2}Nk_BT
\end{equation}
as the average energy per ``bit'' (in certain unit of information
and in terms of the Shannon entropy). Here
\begin{equation}\label{holo}
N=\frac{Ac^3}{G_N\hbar}
\end{equation}
is the number of bits on the boundary with area $A$, as suggested by
the holographic principle. One would be puzzled when checking the
validity of (\ref{equip0}) on the horizon of the RN black hole,
where
\begin{eqnarray}\label{AHTH}
\nonumber A_H&=&\frac{4\pi G_N^2}{c^4}\left(M+\sqrt{M^2-Q^2}\right)^2,\\
T_H&=&\frac{2G_N\hbar}{k_BcA_H}\sqrt{M^2-Q^2}.
\end{eqnarray}
Obviously the relation (\ref{equip0}) is violated as long as
$Q\neq0$. It would be violated even if we go away from the event
horizon of an RN black hole.

Actually this puzzle has driven us to pursue the entropic origin of
Coulomb force. The puzzle will be solved in subsequent sections. To
do this, we will reinterpret the temperature and generalize the
equipartition relation in section \ref{sect-Coulomb}. In the same
section, we will derive Coulomb's law in two different but related
ways. As we will see, there are promising relations between our
postulates and the D-brane picture of black hole thermodynamics. In
section \ref{sect-Maxwell}, taking Coulomb force as an entropic
force, we will derive the Poisson equation and the Maxwell
equations. The holographic principle and the generalized partition
rules are vital in our derivation. Finally, in section
\ref{sect-disc}, we will comment on our results and open questions.

\section{Generalized Equipartition Rule and Coulomb's Law}\label{sect-Coulomb}
Our analysis is very similar to Verlinde's \cite{Verlinde:2010hp}.
We will assume the readers are familiar with the ideas and skills in
reference \cite{Verlinde:2010hp}. In this way, we can pay attention
to our new ingredients and results, while the readers can better
reflect on our ideas.

\subsection{Scheme One}\label{subs-sch1}
There are two schemes to solve the puzzle we brought out. The first
choice is to naively replace the equipartition rule (\ref{equip0})
with
\begin{equation}\label{equip1}
c^2\sqrt{M^2-Q^2}=\frac{1}{2}Nk_BT.
\end{equation}
This relation is well-satisfied on the horizon of the RN black hole.
In the past, we are familiar with the equipartition rule for energy.
When writing down (\ref{equip1}), we generalized the rule to
$\sqrt{M^2-Q^2}$. On the event horizon, the temperature $T$ is
identical to the Bekenstein-Hawking temperature. Outside the
horizon, $T$ can be considered as a generalized Bekenstein-Hawking
temperature on the holographic screen. But such a generalization
makes sense only if $M\geq|Q|$. This is a shortcoming of rule
(\ref{equip1}).

Now suppose there is a test particle with mass $m$ and charge $q$
near a holographic screen which encloses a volume with total mass
$M$ and total charge $Q$. In reference \cite{Verlinde:2010hp},
without a charge, it was shown that the Newton's law emerges as an
entropic force \footnote{We would like to regard (\ref{entrF}) as
the definition of entropic force rather than the first law of
thermodynamics. Some people take it as the first law of
thermodynamics, then the second law of Newton should be implicit,
because $F=\partial_xE=\partial_x(mv^2/2)=mva/v=ma$. If we took
their point of view, then Verlinde's derivation for the second law
of Newton \cite{Verlinde:2010hp} would look like circular
reasoning.}
\begin{equation}\label{entrF}
F=T\partial_xS,
\end{equation}
where $x$ is the emergent coordinate, i.e., the direction
perpendicular to the holographic screen. Generalizing the assumption
of \cite{Verlinde:2010hp}, we write down
\begin{equation}\label{pS1}
\partial_xS=\frac{2\pi k_Bc}{\hbar}\frac{Mm-Qq}{\sqrt{M^2-Q^2}}
\end{equation}
When $Q=0$, it reduces to the assumption made in
\cite{Verlinde:2010hp}, also in the third equation of
(\ref{equippS2}) in the next subsection. In \cite{Verlinde:2010hp}
there is an elegant argument for (\ref{equippS2}), but for
(\ref{pS1}) we have not found such a nice argument. This is another
shortcoming of scheme one.

Nevertheless, by substituting (\ref{holo}), (\ref{equip1}) and
(\ref{pS1}) into (\ref{entrF}), we can unify gravity and Coulomb
force as one entropic force
\begin{equation}\label{entrFgem1}
F=\frac{G_N}{r^2}(Mm-Qq).
\end{equation}

It would be interesting to notice that in the limit $m\ll M$ and
$q\ll Q$, equation (\ref{pS1}) can be reformed as
\begin{equation}
\partial_xS\simeq\frac{2\pi k_Bc}{\hbar}\delta\sqrt{M^2-Q^2}=\frac{\pi k_B^2}{\hbar c}\delta(NT)
\end{equation}
if we identify $\delta M=m$, $\delta Q=q$. This might give us a hint
to incorporate the angular momentum, but we will not go that far in
this paper.

In the present scheme, masses and charges appear in a mixed form, or
namely, in a better-unified form. In the coming subsection, we will
propose a scheme where the mass and the charge are
``disassociated''.

\subsection{Scheme Two}\label{subs-sch2}
The second choice is by postulating equipartition rules and entropy
changes in the emergent direction as following:
\begin{eqnarray}\label{equippS2}
\nonumber Mc^2=\frac{1}{2}Nk_BT_g,&&\partial_xS_g=\frac{2\pi k_Bmc}{\hbar},\\
Qc^2=\frac{1}{2}Nk_BT_{em},&&\partial_xS_{em}=-\frac{2\pi k_Bqc}{\hbar}.
\end{eqnarray}
Remember that, as mentioned in the very beginning of section
\ref{sect-mot}, we are working in the geometrized unit of charge, in
which the Coulomb's law takes almost the same form as the Newton's
law except for the difference in signature. In our assumptions, the
temperature and entropy should be reinterpreted. In this scheme, the
gravitational and the electromagnetic parts have their own
``temperatures'' and ``entropies'', as indicated by the subscripts.

The first line of (\ref{equippS2}) corresponds to the gravitational
part, which has been investigated by Verlinde in
\cite{Verlinde:2010hp}. The new ingredients are these given in the
second line, which are essentially independent of the gravitational
part. This line corresponds to the electromagnetic part. Even the
holographic screen and the emergent direction for electromagnetic
force can be different from those for gravity. Here again we
generalized the equipartition rule, to the electric or magnetic
charge $Q$. The electric or magnetic ``temperature'' $T_{em}$ is
dictated by the average charge per bit. It is important to notice
that $T_{em}$ can be positive or negative, depending on the
signature of $Q$. This sounds bizarre. However, as will be explained
in the next subsection, we can never observe $T_{em}$ directly.
Amusingly, the definition of entropy change in the second line
implies a new interpretation for the electric or magnetic charge,
parallel to Verlinde's interpretation for mass
\cite{Verlinde:2010hp}.

With these assumptions at hand, it is straightforward to derive
Coulomb's law from the entropic force
$F_{em}=T_{em}\partial_xS_{em}$. When the emergent directions of
gravity and electromagnetic force coincide, we can reobtain equation
(\ref{entrFgem1}) as
\begin{equation}\label{entrFgem2}
F_g+F_{em}=T_g\partial_xS_g+T_{em}\partial_xS_{em}=\frac{G_N}{r^2}(Mm-Qq).
\end{equation}

Now return to our puzzle. In view of this scheme, the temperature
$T$ in (\ref{equip0}) should be interpreted as gravitational
temperature $T_g$, and it does not equal to the Bekenstein-Hawking
temperature (\ref{AHTH}) on the horizon. Then the puzzle is solved.

\subsection{Relations between Schemes and to D-brane}\label{subs-relation}
Scheme two has more general applications than scheme one, because
scheme one can be applied only if $M\geq|Q|$ and the distribution of
Newtonian potential coincides with the distribution of Coulomb
potential. It is quite revealing to derive scheme one from scheme
two when these conditions are satisfied. This can be done quickly in
virtue of the relations below:
\begin{eqnarray}
\label{Trel}T^2&=&T_g^2-T_{em}^2,\\
\label{TSrel}T\partial_xS&=&T_g\partial_xS_g-T_{em}\partial_xS_{em}.
\end{eqnarray}

In subsection \ref{subs-sch1}, we mentioned that the temperature in
(\ref{equip1}) is a natural generalization of the Bekenstein-Hawking
entropy outside the event horizon of the RN black hole. From this
point of view, only the temperature $T$ on the left hand side of
(\ref{Trel}) is possibly observable, while $T_{em}$ on the right
hand side is never. $T_g$ is not observable if $T_{em}\neq0$. But
the value of observable $T$ is determined by $T_g$ and $T_{em}$
together.

Up to some numerical coefficients, expression (\ref{AHTH}) holds
well in a higher dimensional spacetime. As a result, in appropriate
units, expressions (\ref{Trel}) and (\ref{TSrel}) will hold in
higher dimensional spacetimes.

Relation (\ref{Trel}) reminds us of the temperatures defined in
reference \cite{Callan:1996dv} in the D-brane picture, where the
thermodynamic property of a 5-dimensional near-extremal RN black
hole has been studied. Here we copy some relations from
\cite{Callan:1996dv} in their convention of notations and
normalization. In that reference, they defined the left-moving
temperature and the right-moving temperature as
\begin{eqnarray}
\nonumber T_L&=&\frac{2}{\pi R}\sqrt{\frac{N_L}{Q_1Q_5}},\\
T_R&=&\frac{2}{\pi R}\sqrt{\frac{N_R}{Q_1Q_5}},
\end{eqnarray}
where $N_L$ is the quantized number of left-moving momentum  and
$N_R$ is the right-moving one. In the case $N_L\gg N_R\gg1$, they
found the Bekenstein-Hawking temperature is identical to the
right-moving temperature $T_H=T_R$.

Stimulated by the D-brane picture, we conjecture that the
temperatures in (\ref{Trel}) can be identified as $T_g=T_L$ and
$T=T_R$. Then making use of some relations in \cite{Callan:1996dv},
we find
\begin{equation}
T_{em}^2=T_L^2-T_R^2=\left(\frac{2}{\pi R}\right)^2\frac{N_L-N_R}{Q_1Q_5}=\left(\frac{2}{\pi r_e}\right)^2,
\end{equation}
in which $r_e$ is horizon radius of the near-extremal black hole.
The same result can be obtained by generalizing (\ref{equippS2}) to
five dimensions.

\section{From Entropic Force to Maxwell's Equations}\label{sect-Maxwell}
\subsection{To Poisson's Equation}
In scheme two of the previous section, we treat the electromagnetic
force and gravity independently. This scheme can be extended to
general matter and charge distributions. As argued in
\cite{Verlinde:2010hp}, the holographic screens correspond
equipotential surfaces, so it seems natural to define the
gravitational temperature generally by the gradient of Newtonian
potential
\begin{equation}
k_BT_g=\frac{\hbar}{2\pi c}\nabla\Phi_g.
\end{equation}
If one generalizes relation (\ref{holo}) to the form
\begin{equation}\label{dholo}
dN=\frac{c^3}{G_N\hbar}dA,
\end{equation}
the Poisson equation for gravity can be perfectly derived
\cite{Verlinde:2010hp}.

Likewise, we define the electromagnetic temperature generally by the
gradient of Coulomb potential
\begin{equation}\label{Tem}
k_BT_{em}=-\frac{\hbar}{2\pi c}\nabla\Phi_{em}.
\end{equation}
Once again we choose the geometrized unit. To recover the familiar
unit, one has to replace
$\Phi_{em}\rightarrow\sqrt{4\pi\varepsilon_0G_N}\Phi_{em}$. Then
making use of the integral expression of the equipartition rule for
charge
\begin{equation}
Qc^2=\frac{1}{2}k_B\oint_{\partial V}T_{em}dN
\end{equation}
and the Gauss' theorem, we straightforwardly work out
\begin{equation}
Q=-\frac{1}{4\pi G_N}\int_V\nabla^2\Phi_{em}dV
\end{equation}
and hence the Poisson equation
\begin{equation}
\nabla^2\Phi_{em}=-4\pi G_N\rho_{em}.
\end{equation}
Transformed from the geometrized unit to the ordinary unit, it takes
the familiar form
\begin{equation}\label{Poisson}
\nabla^2\Phi_{em}=-\frac{1}{\varepsilon_0}\rho_{em}.
\end{equation}

\subsection{To Maxwell's Equations}
With the Poisson equation at hand, the derivation of Maxwell's
equations is quite similar to or even simpler than the derivation of
Einstein's equations in \cite{Verlinde:2010hp}, where Jacobson's
method \cite{Jacobson:1995ab} was followed. In this subsection, we
will sketch the procedure but not go into details.

In order to derive the Maxwell equations, we need a time-like
Killing vector $\xi^a$, just like in \cite{Verlinde:2010hp}. Then
$\rho_{em}$ and $\nabla^2\Phi_{em}$ can be recast into the covariant
form
\begin{eqnarray}
\nonumber \rho_{em}&=&\xi_aj^a,\\
\nabla^2\Phi_{em}&=&\frac{1}{\sqrt{-g}}\xi_a\partial_b\left(\sqrt{-g}F^{ab}\right).
\end{eqnarray}
Subsequently (\ref{Poisson}) is generalized to the covariant form
\begin{equation}
\frac{1}{\sqrt{-g}}\xi_a\partial_b\left(\sqrt{-g}F^{ab}\right)=-\frac{1}{\varepsilon_0}\xi_aj^a.
\end{equation}
Following the trick in \cite{Verlinde:2010hp} and applying
Jacobson's reasoning \cite{Jacobson:1995ab} to time-like screens,
one can obtain the Maxwell equations with a source current $j^a$.

\section{Discussion}\label{sect-disc}
The results in this paper are tentative rather than conclusive. We
will make some comments to help the readers better understand our
results and ponder on them. After a preliminary version of this
paper was posted in the e-Print archive \cite{Wang:2010px}, some
issues mentioned above were later discussed from different
viewpoints in the literature
\cite{Kuang:2010gs,Hossenfelder:2010ih,ChangYoung:2010rz,Li:2010bc}.
The comments below are made after considering the literature.

We are motivated by the RN black hole where $M\geq|Q|$, but most of
the analysis does not rely on the RN solution or the condition
$M\geq|Q|$. Actually, in most parts we treat the gravity and
electromagnetic force independently. Therefore, as long as the
gravitational background is negligible, we can study the pure
electrodynamics with the holographic principle and the generalized
``equipartition rules''. As clarified in \cite{Hossenfelder:2010ih},
the key point is a straightforward generalization from gravitational
``charge'' (mass) to electric charge. The ``puzzle'' proposed in the
beginning of this paper was solved in a better way in
\cite{ChangYoung:2010rz}, but it does not affect the entropic
derivation of electromagnetic force here.

In our formulation, the electromagnetic temperature could be
negative. This looks bizarre. Especially, this impairs the physical
significance of the formally thermodynamical interpretation. At
first sight, one may avoid it by focusing on the systems with the
same signature of charge. If there is no signature difference in
charges, there will be no signature difference in temperatures, and
then one can always choose a positive signature. However, in that
situation we can only study the repulsive Coulomb force. When
physically interpreting our mathematical formulation, the problem of
negative temperature is confronted as a serious obstacle. The
authors in \cite{Li:2010bc} met the same problem when they try to
accommodate inflation in the entropic force scenario. At the moment,
we do not have a good solution to this problem. We hope there will
be progress in the future.

In the above, we investigated the electrostatics, which is trivial
to generalize to incorporate the static magnetic field. Although the
Maxwell equations have been derived based on charge integral, the
time-dependent case has not been studied and would be more
complicated. Of course, the time-dependent gravitational background
is even harder to tame.

In subsection \ref{subs-relation}, we conjectured that our
postulates could have a close relation to the D-brane picture of the
thermodynamics of RN black holes. We conjectured the correspondence
$T_g=T_L$, $T=T_R$. It will be necessary to do more serious and
extensive comparison with the quantities given in
\cite{Callan:1996dv}.

No matter how we formulate the theories and change our logic, the
Nature goes in its own way. It is possible that Newton's law
originates from entropic force but Coulomb's law does not, or
neither does. In the past, we have collected a lot of evidence that
the gravitational force has a holographic nature. Little for the
electromagnetic force. On the other hand, we have a quantum theory
of electrodynamics to explain its microscopic degrees of freedom,
but this is not the case for gravity. Here the motivation is to
reconcile Verlinde's equipartition rule on the horizon of RN black
hole. The other support is given by the apparent similarity between
Newton's law and Coulomb's law. If they really go in the same way,
we hope the formulation in this paper can be regarded as a form of
unification of gravity and electromagnetic force.

\begin{acknowledgments}
This work is supported by the China Postdoctoral Science Foundation.
We are grateful to Miao Li for introducing Verlinde's idea on his
blog.
\end{acknowledgments}

\end{document}